\documentclass[12pt]{article}
\usepackage{fullpage}
\usepackage{latexsym}
\usepackage{amsmath}
\usepackage{amssymb}
\usepackage{amsthm}
\usepackage{amscd}
\usepackage[mathscr]{eucal}

\newcommand{\R}{\mathbb{R}}

\title{A kinetic theory of diffusion in general relativity\\ with cosmological scalar field}

\author{Simone Calogero\footnote{calogero@ugr.es}\\
Departamento de Matem\'atica Aplicada\\ 
Facultad de ciencias, Universidad de Granada\\
18071 Granada, Spain}

\begin{document}
\maketitle

\abstract{
A new model to describe the dynamics of particles undergoing diffusion in general relativity is proposed. The evolution of the particle system is described by a Fokker-Planck equation without friction on the tangent bundle of spacetime. It is shown that the energy-momentum tensor for this matter model is not divergence-free, which makes it inconsistent to couple the Fokker-Planck equation to the Einstein equations. This problem can be solved by postulating the existence of additional matter fields in spacetime or by modifying the Einstein equations. The case of a cosmological scalar field term added to the left hand side of the Einstein equations is  studied in some details. For the simplest cosmological model, namely the flat Robertson-Walker spacetime, it is shown that, depending on the initial value of the cosmological scalar field, which can be identified with the present observed value of the cosmological constant, either unlimited expansion or the formation of a singularity in finite time will occur in the future. Future collapse into a singularity also takes place for a suitable small but positive present value of the cosmological constant, in contrast to the standard diffusion-free scenario. }

\section{Introduction}
Diffusion is one of the most fundamental macroscopic forces in nature. It is the driving mechanism of many dynamical processes in physics, such as heat conduction and  Brownian motion, and it is often invoked to describe transport phenomena in biology or even in social sciences. 
Given the wide variety of phenomena that display diffusive behavior, it is somehow surprising that a consistent theory of diffusion in general relativity is at present still missing. Besides the evident theoretical physics motivation for developing such a theory, diffusion may also play an important role in the large scale evolution of matter in the universe.
% and could help to answer some of the fundamental open questions in cosmology.

The relativistic theory of diffusion processes has received considerable attention in recent years\footnote{See~\cite{AC},~\cite{angfra}--\cite{baifra},~\cite{dmr}--\cite{DTH},~\cite{frajan}--\cite{herr2}.  The review~\cite{DH2} contains more references and an historical introduction to the relativistic theory of diffusion.}. Most works so far have been concerned with the justification and the analysis of the stochastic differential equations that describe diffusion at the microscopic level. The theory started with the pioneering work of Dudley~\cite{dudley}, who showed that Lorentz invariant Markov processes do not exist in Minkowski space, while only one such process with continuous paths (Brownian motion) can be defined on the tangent bundle (the relativistic phase space). Dudley's process was generalized to curved manifolds in~\cite{frajan} and studied on some exact manifold solutions of the Einstein equations of general relativity in~\cite{angst,angst2,frajan2,haba2,herr2}. These works however are only concerned with describing the diffusion of particles {\it on a given background spacetime}, and no attempt is made to relate the curvature of spacetime with the dynamics of the matter undergoing diffusion. 

A different approach to diffusion is based on the kinetic Fokker-Planck equation, which is the partial differential equation satisfied by the (one-particle) distribution function $f$. Although the Fokker-Planck equation is formally equivalent (via It\^o's formula) to a system of stochastic differential equations, it is often easier to justify the evolution of the function $f$ than the dynamics of the microscopic stochastic process (for instance by an approximation procedure starting from a master or Boltzmann equation~\cite{landau,Risken}). This point of view was taken in~\cite{AC}, where a relativistic Fokker-Planck equation previously derived in~\cite{DH1} by stochastic calculus methods, was re-discovered as the most natural second order transport-diffusive equation that is Lorentz invariant in the absence of friction\footnote{There exist other models in the literature that are named ``relativistic Fokker-Planck equations''~\cite{dmr,DH1}, but none of them is Lorentz invariant in the absence of friction.}. Moreover, the associated stochastic differential equations, when expressed in terms of the proper time, reduce to Dudley's process~\cite{herr}.

When considering diffusion in general relativity, the need to rely on the macroscopic continuum description provided by the Fokker-Planck equation becomes transparent. In general relativity, in fact, the geometry of spacetime is not given in advance but it is determined by the Einstein equations,
\begin{equation}\label{Einsteinequation}
R_{\mu\nu}-\frac{1}{2}g_{\mu\nu}R=T_{\mu\nu},\quad (8\pi G=c=1),
\end{equation}
where the matter fields enter through the energy-momentum tensor $T_{\mu\nu}$. In kinetic theory the energy-momentum tensor  is given by a suitable integral of the particle distribution $f$, which in the presence of diffusion is the solution of the Fokker-Planck equation. In its turn, the metric, i.e., the solution of the Einstein equations, appears in the Fokker-Planck equation. Thus the consistent description of matter undergoing diffusion in general relativity requires to study the coupled system of Einstein and Fokker-Planck equations. 

However it is not difficult to see that a matter distribution undergoing diffusion cannot appear as the only source in the Einstein equations~\eqref{Einsteinequation}. The reason is that under the action of the diffusion forces the kinetic energy of the particles is not preserved, and accordingly the energy-momentum tensor $T^{\mu\nu}$ does not fulfill the compatibility condition $\nabla_\mu T^{\mu\nu}=0$ required by the Bianchi identities and the Einstein equations. A possible solution to this problem is  to assume the existence of additional matter fields in spacetime which exchange energy with the particles undergoing diffusion. Another solution, which is studied in some details in the present paper, is to add a cosmological scalar field term in the left hand side of~\eqref{Einsteinequation}, leading to the following modification of Einstein's equations
\begin{equation}\label{Einsteinequation2}
R_{\mu\nu}-\frac{1}{2}g_{\mu\nu}R+\phi g_{\mu\nu}=T_{\mu\nu}.
\end{equation}
The evolution of the scalar field $\phi$ is determined by the particle distribution function $f$ through the equation $\nabla_\nu\phi=\nabla^\mu T_{\mu\nu}$, which follows by~\eqref{Einsteinequation2} and the Bianchi identity $\nabla^\mu (R_{\mu\nu}-\frac{1}{2}g_{\mu\nu}R)=0$. In fact, it will be shown that the energy-momentum tensor associated to a solution of the Fokker-Planck equation satisfies
\[
\nabla_\mu T^{\mu\nu}=3\sigma J^\nu,
\]
where $\sigma>0$ is the diffusion constant and $J^\mu$ is the current density of the matter. Thus the dynamics of the scalar field $\phi$ is ruled by the equation
\begin{equation}\label{eqpsiintro}
\nabla_\mu\phi=3\sigma J_\mu.
\end{equation} 
In vacuum (i.e., $T^{\mu\nu}= J^\mu=0$), or in the absence of diffusion ($\sigma=0$),  $\phi$ is constant throughout spacetime and the evolution equations~\eqref{Einsteinequation2} reduce to the Einstein equations with cosmological constant. When diffusion is present, we get a ``variable cosmological constant'' model. Note however that in contrast to ``the variable cosmological constant'' theories considered in the literature\footnote{See~\cite{arxive} for a review.}, in the present situation the dynamics of the cosmological term is not prescribed {\it a priori}, but it is determined by the matter fields through the equation~\eqref{eqpsiintro}. 

The general relativistic theory of diffusion briefly outlined above is developed in more details in sections~\ref{curvedFP}-\ref{coupling} and it is applied in section~\ref{robwal} to the cosmological model with the simplest geometry, namely the flat Robertson-Walker spacetime. This model is simple enough to avoid serious mathematical difficulties and at the same time it is sufficient to derive some interesting physical implications of the new theory. In particular, it will be shown that the model predicts future collapse of the universe into a singularity in finite time even for positive (but sufficiently small) initial values $\phi_0$ of the cosmological scalar field. Since $\phi_0$ can be identified with the current observed value of the cosmological constant  $\Lambda$, this result is in contrast with the conclusions of the standard (diffusion-free) cosmological model with $\Lambda>0$, which predicts an unlimited phase of accelerated expansion of the universe in the future. The reason for the formation of a singularity in the presence of diffusion is that the cosmological scalar field $\phi$, although initially positive, may become negative at later times, after which the cosmological model behaves qualitatively like a solution of the Einstein equations with negative cosmological constant. 
%Moreover the cosmological scalar field $\phi$ is a decreasing function of the cosmological time $t$, thus it could have had a very large value in the early universe, in accordance with the predictions of particle physics~\cite[pag.~16]{rich}.  

Before going to the main topic of the paper, I shall review in the next section some aspects of the kinetic theory of diffusion in the non-relativistic and special relativistic case. 

%PUT LATER
%There is another important difference between the FP equation considered here and the one studied in Debbasch,  Hermmann (haba, franchi?). When the effect of disffusion in our equation is neglected, i.e, when we set $\sigma=0$ in the equation, where $\sigma$ is the diffusion constant, the particle distribution describes an ensemble of particles moving along the geodesics of the spacetime. Since in the absence of diffusion, gravity is the only remained interaction among the particles (i.e., the particles are  free falling), the fact that they have to move along geodesics is in accordance with the equivalence principle. In particular, when the diffusion term is removed, our equation reduces to the Liouville (or Vlasov) equation.  

%Finally, we neglect friction for the following reasons: first it is not clear, on physical grounds, how the friction term should look like; secondly, friction, as opposed to diffusion, refers to a particular reference frame (e.g., the rest frame of a termal bath). Thus in order to introduce friction in general relativity, the existence of a privilegied observer should be assumed, which is against the geometric and philosophical interpretation of general relativity.  
 
\section{Kinetic diffusion on flat spacetimes}\label{kindiff}
Let $f(t,x,p)\geq 0$ denote the distribution function in phase space of an ensemble of unit mass particles. The integral of $f(t,x,p)$ in a region $V\times U$ of phase space is the number of particles which, at the time $t\geq0$, have position $x\in V$ and momentum $p\in U$. The total number of particles in the system is then
\[
n(t)=\int_{\R^3}\int_{\R^3}f(t,x,p)dpdx.
\]   
The kinetic evolution of the particles system is governed by a partial differential equation on $f$; the particular form of this equation depends on the interaction among the particles. 
%Two standard conditions that any such PDE should satisfy are: (i) preserving the non-negativity of $f$ (i.e., if $f(0,x,p)\geq 0$, then $f(t,x,p)\geq 0$); (ii) preserving the total number of particles: $N=const.$ 
The most basic example is the non-relativistic free-transport equation:
\begin{equation}\label{freenonrel}
\partial_tf+p\cdot\nabla_xf=0.
\end{equation}   
The free-transport equation describes the kinetic evolution of free-moving particles. If the particles are interacting either by internal or external forces, a new term is to be added in the right hand side of~\eqref{freenonrel}. For instance
\begin{equation}\label{FPnonrelnofr}
\partial_tf+p\cdot\nabla_xf=\sigma\Delta_pf,
\end{equation} 
is the kinetic Fokker-Planck (or Kramers) equation in the absence of friction~\cite{Risken}. Here $\sigma>0$ is the diffusion constant and $\Delta_p$ denotes the Laplace operator in the $p$ variable. The new term $\Delta_pf$ is the most common and basic way to model {\it diffusion} in the momentum variable. The main physical assumption behind equation~\eqref{FPnonrelnofr} is that the particles are moving in a background fluid in thermal equilibrium (thermal bath).  Assuming that the molecules of the fluid are much lighter than the particles, and that the particles make up a sufficiently dilute system, the total force acting on the particles can be macroscopically approximated by two dominant contributions: diffusion, which is due to thermal fluctuations and is therefore associated---at the microscopic level---to random collisions with the molecules of the fluid, and {\it friction}, which takes into account deterministic grazing collisions among the particles. In this paper the contribution due to friction is neglected.

%The general solution of~\eqref{FPnonrelnofr} with initial data $f_0$ can be written as
%\begin{equation}\label{gensolFPnonrelnofr}
%f(t,x,p)=\int_{\R^3}\int_{\R^3}G(t,x',p';x,p)f_0(x',p')dx'dp',\quad t>0,
%\end{equation}
%where $G$ is the fundamental solution of~\eqref{FPnonrelnofr}, whose exact form can be found for instance in. 
Note that $n=const.$, i.e., the total number of particles is conserved. This is a trivial consequence of the divergence theorem:
\[
\frac{dn}{dt}=\int_{\R^3}\int_{\R^3}\partial_t f dp dx=\int_{\R^3}\int_{\R^3}(-p\cdot\nabla_xf+\sigma\Delta_pf)dpdx=0.
\]
Furthermore the average kinetic energy, defined as
\[
\mathcal{E}_{\mathrm{kin}}(t)=\frac{1}{2}\int_{\R^3}|p|^2fdpdx,
\]
increases linearly in time\footnote{In particular, the particle system is unable to relax to an equilibrium state under the action of the diffusion forces alone. To this purpose, a friction term is to be added to the right hand side of~\eqref{FPnonrelnofr}, see~\cite{Risken}.}, since
\begin{equation}\label{ekingrow}
\frac{d}{dt}\mathcal{E}_{\mathrm{kin}}(t)=3\sigma n.
\end{equation}

The relativistic, Lorentz invariant generalization of~\eqref{FPnonrelnofr} is\footnote{Throughout the paper, greek indexes run from 0 to 3, latin indexes from 1 to 3 and the Einstein summation rule applies. Moreover, physical units are fixed such that $8\pi G=c=1$, where $G$ is Newton's gravitational constant and $c$ is the speed of light.}:
\begin{equation}\label{FPrelnofr}
p^\mu\partial_{x^\mu}f=\sigma\Delta^{(h)}_pf,
\end{equation}
where $p^{\mu}$---the particles 4-momentum---satisfies the {\it mass shell} condition 
\[
\eta_{\mu\nu}p^{\mu}p^\mu=-1,
\]
where $\eta$ is the Minkowski metric. The previous equation  can be used to express the time component of the 4-momentum in terms of its spatial components: 
\[
p^0=\sqrt{1+|p|^2},\quad p=(p^1,p^2,p^3). 
\]
The operator in the left hand side of~\eqref{FPrelnofr} is the relativistic free-transport operator. In the right hand side,
$\Delta^{(h)}_p$ denotes the Laplace-Beltrami operator associated to the hyperbolic metric $h$: 
\[
h_{ij}=\delta_{ij}-\hat{p}_i\hat{p}_j,
\]
where $\hat{p}=p/p^0$ is the relativistic velocity. Note that $h$ is the Riemannian metric induced by the Minkowski metric over the hyperboloid $p^0=\sqrt{1+|p|^2}$. For a justification of~\eqref{FPrelnofr} see~\cite{AC}, where the more general case with friction is discussed. The explicit form of $\Delta^{(h)}_pf$ is 
\[
\Delta^{(h)}_pf=\frac{1}{\sqrt{\det h}}\partial_{p^i}\left(\sqrt{\det h}\,(h^{-1})^{ij}\partial_{p^j}f\right)=\frac{1}{p^0}\partial_{p^i}\left(\frac{\delta^{ij}+p^ip^j}{p^0}\partial_{p^j}f\right),
\]
where $\det h=1+|p|^2$, and $(h^{-1})^{ij}=\delta^{ij}+p^ip^j$ is the inverse matrix of $h_{ij}$. It follows that~\eqref{FPrelnofr} can be written in the divergence form
\begin{equation}\label{FPrelnofr2}
\partial_t f+\hat{p}\cdot\nabla_x f=\sigma\partial_{p^i}\left(\frac{\delta^{ij}+p^ip^j}{p^0}\partial_{p^j}f\right),
\end{equation}
by which it is clear that $n=const$. Moreover, defining the average kinetic energy of the particles as
\[
\mathcal{E}_\mathrm{kin}(t)=\int_{\R^3}\sqrt{1+|p|^2}fdpdx,
\]
it is easy to check that equation~\eqref{ekingrow} holds in the relativistic case as well.
%Also the general solution of~\eqref{FPrelnofr} can be written in the form~\eqref{gensolFPnonrelnofr}, but the explicit form of the fundamental solution in this case is not known.

The relativistic current density vector and energy-momentum tensor are given by
\begin{align*}
&J^\mu(t,x)=\int_{\R^3}f(t,x,p)\, p^\mu \frac{dp}{p^0},\\
&T^{\mu\nu}(t,x)=\int_{\R^3}f(t,x,p)\, p^\mu p^\nu\frac{dp}{p^0},
\end{align*}
independently of the equation satisfied by the distribution function $f$.
A straightforward calculation using~\eqref{FPrelnofr2} leads to the identities
\begin{equation}\label{divT}
\partial_{x^\mu}J^\mu=0,\quad
\partial_{x^\mu}T^{\mu\nu}=3\sigma J^\nu,
\end{equation}
which imply
\begin{equation}\label{div2T}
\partial_{x^\mu}\partial_{x^\nu}T^{\mu\nu}=0.
\end{equation}
%The fact that $T_{\mu\nu}$ is not divergence free means that the relativistic Fokker-Planck equation does not conserve energy. This is of course not surprising: the non-relativistic model~\eqref{FPnonrelnofr} does not preserve energy either. 

To conclude this section, let us comment on the fact that the particles kinetic energy is not conserved in the models considered above.  The physical explanation for this fact is that, while the particles are colliding with the molecules of the surrounding fluid, the change of energy of the fluid molecules is neglected in the diffusion approximation, since the fluid is assumed to be in thermal equilibrium. If the dynamics of the coupled system particles-fluid were considered, then the total energy would be conserved. However, as shown in this section, it is possible to consider an approximation of the full dynamics by looking at the evolution of the particle distribution $f$ only and still end up with a consistent equation for $f$, which is Galilean invariant in the non-relativistic case (eq.~\eqref{FPnonrelnofr}) and Lorentz invariant in the special relativistic case (eq.~\eqref{FPrelnofr}). We shall see that in general relativity this approximation is inconsistent.

\section{Kinetic diffusion on curved spacetimes}\label{curvedFP}
Suppose now that the background Minkowski spacetime is replaced by a general Lorentzian, time-oriented manifold $(M,g)$. For the moment the metric $g$ is assumed to be given. Let $x$ denote an arbitrary point of $M$ and $x^\mu$ a (local) system of coordinates on an open set $U\subset M$, $x\in U$, with $x^0\equiv t$ being timelike. The vectors $\partial_{x^\mu}$ form a basis of the tangent space $T_xM$ and  the components of $p\in T_xM$ in this basis will be denoted by $p^\mu$. $(x^\mu,p^\nu)$ provides a system of coordinates on $TU\subset TM$, where $TM$ denotes the tangent bundle of $M$. The (future) mass-shell is the 7-dimensional submanifold of the tangent bundle defined as
\[
\Pi M=\{(x,p)\in TM: g(x)(p,p)=-1,\, p\text{ future directed}\}.
\]
On the subset $\Pi U=\{(x,p)\in\Pi M:x\in U\}$ of the mass-shell, the condition $g(x)(p,p)=-1$ is equivalent to $g_{\mu\nu}p^\mu p^\nu =-1$ (where $g_{\mu\nu}=g_{\mu\nu}(x^\alpha)$), which can be used to express $p^0$ in terms of $p^1,p^2,p^3$, precisely:
\[
p^0=-\frac{1}{g_{00}}\left[g_{0j}p^j+\sqrt{(g_{0j}p^j)^2-g_{00}(1+g_{ij}p^ip^j)}\right],
\]
where the choice of the positive root reflects the condition that $p$ is future directed. 
Differentiating the mass shell relation $g_{\mu\nu}p^\mu p^\nu=-1$ one obtains the useful relations
\begin{equation}\label{useful}
\partial_{x^\mu}p^0=-\frac{p^\alpha p_\beta}{p^0}\Gamma^\beta_{\ \mu\alpha},\quad \partial_{p^j}p^0=-\frac{p_j}{p^0},
\end{equation}
which will be used below to derive some important identities. In the previous equations, $\Gamma^\beta_{\ \mu\nu}$ denote the Christoffel symbols of $g$ and the indexes are lowered and raised with the matrix $g_{\mu\nu}$ and its inverse $g^{\mu\nu}$.  

Let $\widetilde{L}$ denote the geodesic flow vector field on the tangent bundle. In the vector fields basis $(\partial_{x^\mu},\partial_{p^\nu})$ it is given by
\[
\widetilde{L}=p^\mu(\partial_{x^\mu}-\Gamma^\nu_{\ \mu\alpha }p^\alpha\partial_{p^\nu}).
\]
The Liouville, or Vlasov, operator $L$ is defined as the projection of $\widetilde{L}$ on the mass shell. Using~\eqref{useful} it is easy to derive the following local coordinates representation of $L$:
\[
L=p^\mu(\partial_{x^\mu}-\Gamma^i_{\ \mu\alpha }p^\alpha\partial_{p^i}).
\]
The first fundamental change compared to the flat case is the definition of the ``free-transport operator''. It is now assumed, in agreement with the equivalence principle, that in the absence of any interaction other than gravity, i.e., in free-falling motion, the particles move along the geodesics of $(M,g)$. So now the free-transport equation~\eqref{freenonrel} is replaced by the Vlasov equation:
\begin{equation}\label{vlasov}
L f=0,\quad\text{i.e.,}\quad p^\mu\partial_{x^\mu}f-\Gamma^i_{\ \mu\nu}p^\mu p^\nu\partial_{p^i}f=0,
\end{equation}
where $f:\Pi M\to [0,\infty)$ is the distribution function of particles, which is a smooth function on the mass-shell.
See~\cite{And, ehlers, rendall} for an introduction to kinetic theory and the Vlasov equation in general relativity. Physical systems which are supposed to be well-modeled by the Vlasov equation include galaxies, in which stars play the role of the particles, or even clusters of galaxies, where the galaxies themselves are identified with the Vlasov particles~\cite{galacticdynamics}. Kinetic theory and the Vlasov equation also have important applications in cosmology~\cite{vlasovIX}. 

To transform~\eqref{vlasov} into a Fokker-Planck equation we need to add a diffusion operator on the right hand side. Let $\Pi_xM$ denote the fiber over $x\in M$ of the mass-shell and $\pi_x:\Pi M\to\Pi_x M$ the canonical projection onto it. The action of the diffusion operator on $f$ will be defined as a differential operator acting on $f\circ\pi_x$. The quadratic form $g(x)$ induces a Riemannian metric $h(x)$ on $\Pi_xM$. The components of $h$ are functions of the coordinates $(x^\mu,p^i)$, which, by~\eqref{useful}, are given by
\begin{equation}\label{hij}
h_{ij}=g_{ij}-\frac{p_i}{p^0}g_{0j}-\frac{p_j}{p^0}g_{0i}+g_{00}\frac{p_ip_j}{(p^0)^2}.
\end{equation}
In analogy with the diffusion operator on the mass-shell of Minkowski space defined in section~\ref{kindiff}, we now define the action of the diffusion operator on $f$ by the formula
\[
\mathcal{D}_pf=\Delta^{(h)}_p(f\circ\pi_x),
\]
where $\Delta^{(h)}_p$ is the Laplace-Beltrami operator of the Riemannian metric $h$ on $\Pi_xM$. The subscript $p$ reminds that the operator $\mathcal{D}_p$ acts on the tangent space variables only. The expression of $\mathcal{D}_p$ in local coordinates is given as before by 
\[
\mathcal{D}_pf=\frac{1}{\sqrt{\det h}}\partial_{p^i}\left(\sqrt{\det h}\,(h^{-1})^{ij}\partial_{p^j}f\right),
\]
where now $h_{ij}=h_{ij}(x^\alpha,p^k)$ is the matrix~\eqref{hij}.
The generalization of~\eqref{FPrelnofr} on the curved spacetime $(M,g)$ is then given by
\begin{equation}\label{FPgenrelnofr}
p^\mu\partial_{x^\mu}f-\Gamma^i_{\ \mu\nu}p^\mu p^\nu\partial_{p^i}f=\sigma\mathcal{D}_p f.
\end{equation} 
The current density $J^\mu$ and the energy-momentum tensor $T^{\mu\nu}$ are defined as
\begin{equation}
J^\mu(x)=\sqrt{|g|}\int_{\Pi_xU}f\: \frac{p^\mu}{-p_0}dp^{123},
\end{equation}
\begin{equation}
T^{\mu\nu}(x)=\sqrt{|g|}\int_{\Pi_xU}f\: \frac{p^\mu p^\nu}{-p_0}dp^{123}. \label{TFP}
\end{equation}
In the previous definitions, $|g|=|\det g|$ and $dp^{123}$ denotes the 1-form $dp^1\wedge dp^2\wedge dp^3$. Since $-\sqrt{|g|}/p_0=\sqrt{\det h}$, $T^{\mu\nu}$ and $J^\nu$ are, as they should be, integrals of $f$ on the mass-shell with respect to the invariant measure induced thereon. Note also that, by the mass shell condition $g_{\mu\nu}p^\mu p^\nu=-1$,
\begin{equation}\label{trT}
\mathrm{Tr}_gT=g_{\mu\nu}T^{\mu\nu}\leq 0.
\end{equation}
(Equality holds for particles with zero rest mass.)
Moreover it is shown in~\cite{ehlers} that $J^\mu$ is a timelike vector field and that $T^{\mu\nu}$ verifies the dominant and strong energy conditions.

Now denote by $\nabla$ the Levi Civita connection of the metric $g$. The analogs of the identities~\eqref{divT} hold:
\begin{align}
\nabla_\mu J^\mu&=0,\label{divNgen}\\
\nabla_\mu T^{\mu\nu}&=3\sigma J^\nu.\label{divTgen}
\end{align}
The easiest way to establish the previous identities at the arbitrary point $x\in U$ is to use a coordinates system $\dot{x}^\mu$ such that $g_{\mu\nu}=\eta_{\mu\nu}$ and $\partial_{x^\alpha}g_{\mu\nu}=0$ at $x$. For instance, for~\eqref{divTgen} we have
\begin{align*}
\nabla_\mu T^{\mu\nu}&=\ (\nabla_\mu\log\sqrt{|g|})T^{\mu\nu}+\sqrt{|g|}\int_{\Pi_xU} p^\nu p^\mu\nabla_\mu f\,\frac{dp^{123}}{-p^0}
+\sqrt{|g|}\int_{\Pi_xU} f\,\nabla_\mu\left(\frac{p^\mu p^\nu}{-p_ 0}\right)dp^{123}\\
%=&(\nabla_\mu\log\sqrt{|g|})T^{\mu\nu}
%+\sqrt{|g|}\int_{\R^3}p^\nu\Gamma^i_{\ \alpha\beta}p^\alpha p^\beta\partial_{p^i}f\,\frac{dp^{123}}{-p^0}\\
%&+\sqrt{|g|}\int_{\R^3}f\,\nabla_\mu\left(\frac{p^\mu p^\nu}{-p_ 0}\right)dp^{123}
%+\sqrt{|g|}\,\sigma\int_{\R^3}p^\nu\mathcal{D}_pf\,\frac{dp^{123}}{-p_0}\\
&\doteq\ \sigma\int_{\Pi_xU}p^\nu\partial_{p^i}\left(\frac{\delta^{ij}+p^ip^j}{\sqrt{1+|p|^2}}\right)\partial_{p^j}f\,dp^{123} ,
\end{align*}
where the symbol $\doteq$ means that the equality holds at $x\in U$ in the coordinates $\dot{x}^\mu$. Integrating by parts in the last integral we conclude that~\eqref{divTgen} holds at $x$ in the coordinates $\dot{x}^\mu$ and therefore in any other coordinates. The proof of~\eqref{divNgen} is similar. 
Finally by~\eqref{divNgen}-\eqref{divTgen} the analog of equation~\eqref{div2T} holds
\begin{equation}\label{div2Tgen}
\nabla_\mu\nabla_\nu T^{\mu\nu}=0.
\end{equation}

\section{Coupling with the Einstein equations}\label{coupling}
In this section I will address the main question of this paper, which is how to couple the Fokker-Planck equation to the Einstein equations
\begin{equation}\label{einstein}
R_{\mu\nu}-\frac{1}{2}R\,g_{\mu\nu}=T_{\mu\nu}.
\end{equation}
By the Bianchi identity $\nabla^\mu(R_{\mu\nu}-\frac{1}{2}g_{\mu\nu}R)=0$, the energy momentum tensor must satisfy $\nabla^\mu T_{\mu\nu}=0$, otherwise solutions of the Einstein-matter system cannot exist. Since for solutions of the Fokker-Planck equation~\eqref{FPgenrelnofr} the energy-momentum tensor~\eqref{TFP} is not divergence free, it is then meaningless to use the tensor~\eqref{TFP} in the right hand side of~\eqref{einstein}. This incompatibility can be solved by assuming that there exist other matter fields in spacetime\footnote{The same hypothesis is introduced in the so-called gravitational aether theories, see~\cite{kamiab} and the references therein.}. The role of these additional matter fields is similar to that of the thermal bath in the theory outlined in section~\ref{kindiff}. Let $\mathscr{T}_{\mu\nu}$ denote the energy-momentum tensor of the additional matter fields and consider the Einstein equations
\begin{equation}\label{einstein2}
R_{\mu\nu}-\frac{1}{2}R\,g_{\mu\nu}=T_{\mu\nu}+\mathscr{T}_{\mu\nu}.
\end{equation}
Now the Bianchi identities imply, using ~\eqref{divTgen},
\begin{equation}\label{mattereqs}
\nabla_\mu \mathscr{T}^{\mu\nu}=-3\sigma J^\nu.
\end{equation}
As a way of example, consider a system of particles undergoing diffusion in a perfect fluid. In this case the tensor $\mathscr{T}_{\mu\nu}$ is given by
\[
\mathscr{T}_{\mu\nu}=(\rho+p)u_\mu u_\nu+p\, g_{\mu\nu},
\]
where $\rho$ is the rest frame energy density, $p$ the pressure and $u^\mu$ the 4-velocity of the fluid. Projecting~\eqref{mattereqs} in the direction of $u^\mu$ and onto the plane orthogonal to $u^\mu$ we obtain 
\begin{align}
&\nabla_\mu(\rho u^\mu)+p\nabla_\mu u^\mu=3\sigma  J^\mu u_\mu,\\
&(\rho + p)u^\mu\nabla_\mu u_\nu
+u_\nu u^\mu\nabla_\mu p+\nabla_\nu p=-3\sigma \big(J_\nu+u_\nu(J_\mu u^\mu)\big).
\end{align}
For $\sigma=0$ (i.e., when the particles and the fluid are not interacting), the previous equations reduce, respectively, to the continuity equation and the Euler equation of a perfect fluid in general relativity, see~\cite{wald}.

An alternative solution to the incompatibility of equations~\eqref{FPgenrelnofr} and~\eqref{einstein} is to modify the left hand side of the Einstein equations by adding to it a new tensor term $\mathscr{K}_{\mu\nu}$. Although this solution is formally equivalent to the previous one (by setting $\mathscr{T}_{\mu\nu}=-\mathscr{K}_{\mu\nu})$, it is important to distinguish their physical meaning. The simplest choice for $\mathscr{K}_{\mu\nu}$ is the {\it cosmological scalar field term}
\begin{equation}\label{varcosmoT}
\mathscr{K}_{\mu\nu}=\phi g_{\mu\nu},
\end{equation}
where $\phi$ is a scalar field\footnote{In quantum fields theory it is customary to interpret $\phi$ as the vacuum energy.}. This choice leads to the following modification of~\eqref{einstein}:
\[
R_{\mu\nu}-\frac{1}{2}R\,g_{\mu\nu}+\phi g_{\mu\nu}=T_{\mu\nu}.
\]   
The evolution equation for $\phi$ resulting from~\eqref{divTgen} is 
\begin{equation}\label{mattereqpsi}
\nabla_\mu\phi=3\sigma J_\mu.
\end{equation}
In the absence of diffusion, $\phi$ is constant. In this case the model reduces to the Einstein-Vlasov system with cosmological constant. 

In the rest of the paper we shall focus on the cosmological scalar field model, since it provides a simple and yet physically interesting framework to study the effects of diffusion on the dynamics of spacetime and matter.  
%\footnote{Actually, because of~\eqref{waveq}, only the initial data for the matter enter explicitly in the dynamics of $\phi$; see also the example in the next section.}. 
%In the points of spacetime where $\phi>0$, the scalar field behaves like a positive cosmological constant, i.e., like repulsive gravitational forces, while in the points where $\phi<0$, it enhances gravity, and therefore behaves like a negative cosmological constant.

By~\eqref{mattereqpsi},  diffusion in a scalar field is only possible when $J^\mu$ is irrotational:
\begin{equation}\label{irrotational}
\nabla_\mu J_\nu-\nabla_\nu J_\mu=0.
\end{equation}
The previous equation imposes a rather severe restriction on the solution $f$ of the Fokker-Planck equation~\eqref{FPgenrelnofr}, which makes the question of existence of solutions a highly non-trivial task. An important example where~\eqref{irrotational} holds is the class of spatially homogeneous and isotropic spacetimes, which are the most popular models for the universe in cosmology. Cosmological models in this class are studied in the next section.

By defining the (average) 4-velocity of the matter as 
\[
u^\mu=\frac{J^\mu}{\sqrt{-J^\alpha J_\alpha}},
\]
and using~\eqref{mattereqpsi} we obtain
\begin{equation}\label{decreasing}
u^\mu\nabla_\mu\phi=-3\sigma (-J^\mu J_\mu)^{1/2}.
\end{equation} 
It follows by~\eqref{decreasing} that the cosmological scalar field $\phi$ is decreasing along the matter flow. Hence the model under discussion predicts that the particles gain energy by diffusion and that this energy is provided by the cosmological scalar field $\phi$.
%, i.e., that the energy $(-\phi)$ of the cosmological scalar field (as measured in a frame adapted to $u^\mu$) is increasing. 
Moreover it follows by~\eqref{div2Tgen} that the cosmological scalar field $\phi$ satisfies the homogeneous wave equation:
\begin{equation}\label{waveq}
\Box\phi=0,\quad \Box=\nabla^\mu\nabla_\mu.
\end{equation}
This implies in particular that the energy-momentum tensor of $\phi$,
\begin{equation}\label{diften}
S_{\mu\nu}=\nabla_\mu\phi\nabla_\nu\phi-\frac{1}{2}g_{\mu\nu}\nabla^\alpha\phi\nabla_\alpha\phi
\end{equation} 
is divergence free: 
\begin{equation}\label{divS}
\nabla^\nu S_{\mu\nu}=0.
\end{equation}
Therefore the cosmological scalar field propagates throughout spacetime in form of waves without dissipation.

%Using~\eqref{mattereqpsi}, the tensor $S_{\mu\nu}$ may be rewritten as
%\[
%S_{\mu\nu}=9\sigma^2(J_\mu J_\nu-\frac{1}{2}g_{\mu\nu}J^\alpha J_\alpha ).
%\]

\section{The spatially homogeneous and flat isotropic cosmological model}\label{robwal}  
The purpose of this section is to study the qualitative behavior of spatially homogeneous and isotropic solutions of the system
\begin{align}
&R_{\mu\nu}-\frac{1}{2}R\,g_{\mu\nu}+\phi g_{\mu\nu}=T_{\mu\nu}, \label{Einstein}\\
&p^\mu\partial_{x^\mu}f-\Gamma^i_{\mu\nu}p^\mu p^\nu\partial_{p^i}f=\sigma\mathcal{D}_p f,\label{FokkerPlanck}
\end{align}
where $T_{\mu\nu}$ is given by~\eqref{TFP}. In this application, the particles undergoing diffusion represent the galaxies. 

The symmetry assumption means two things. Firstly that there exist six Killing vectors fields $X_A$, $A=1,\dots 6$ on spacetime, whose orbits are spacelike hypersurfaces. 
Secondly that the distribution function $f$ satisfies $\tilde{X}_Af=0$, where $\tilde{X}_A$, $A=1,\dots 6$ are the lifts over the mass shell of the Killing vectors $X_A$. Under the given symmetry assumptions, the metric can be written in the standard form
\[
ds^2=-dt^2+a(t)^2\left[\frac{dr^2}{1-k r^2}+r^2(d\theta^2+\sin^2\theta d\phi^2)\right],
\]
where $k=0$ or $\pm 1$ is the curvature parameter of the hypersurfaces $t=const.$ 
As proved in~\cite{mama}, the symmetry assumption on the distribution function restricts $f$ to be of the form $f=F(t,q)$, where $q=(p_1^2+p_2^2+p_3^2)^{1/2}$. In particular, the 4-current $J^\mu$ and the energy momentum tensor $T_{\mu\nu}$ take the form
\[
J^\mu=(n,0,0,0),\quad T_{\mu\nu}=\mathrm{diag}(\rho,\mathscr{P},\mathscr{P},\mathscr{P}),
\]
where the total number of particles $n$ is given by
\[
n(t)=\frac{4\pi}{a(t)^3}\int_0^\infty F(t,q) q^2dq,
\]
while the energy density $\rho$ and the pressure $\mathscr{P}$ are given by
\begin{align}\label{matter}
&\rho(t)=\frac{4\pi}{a(t)^4}\int_0^\infty q^2F(t,q) \sqrt{a(t)^2+q^2}dq,\\
\nonumber\\
&\mathscr{P}(t)=\frac{4\pi}{3a(t)^4}\int_0^\infty \frac{q^4F(t,q)}{ \sqrt{a(t)^2+q^2}}dq.
\end{align}
Observe the inequalities
\[
\rho(t)\geq n(t),\quad\mathscr{P}(t)\leq\frac{1}{3}\rho(t),
\]
the second one being equivalent to~\eqref{trT} under the assumptions of spatial homogeneity and isotropy.  
To write the Fokker-Planck equation it is more convenient to use the coordinates $p_1,p_2,p_3$, rather than $q$. Setting $F(t,q)=f(t,p_1,p_2,p_3)$, eq.~\eqref{FokkerPlanck} simplifies to 
\begin{equation}\label{FPequation}
\partial_tf=\partial_{p_ i}\left[\left(\frac{a(t)^2\delta_{ij}+p_ip_j}{p^0}\right)\partial_{p_j}f\right],
\end{equation}
where
\[
p^0=\sqrt{1+\frac{p_1^2+p_2^2+p_3^3}{a(t)^2}}.
\]
Note that in the Minkowski case ($a\equiv 1$),~\eqref{FPequation} reduces to~\eqref{FPrelnofr2} in the spatially homogeneous case. 

The cosmological scalar field is a function of $t$ only and  the equation~\eqref{mattereqpsi} becomes
\[
\dot{\phi}=-3\sigma\,n.
\]
The local conservation of the number of particles, eq.~\eqref{divNgen}, implies
\[
n(t)=\frac{n_0a_0^3}{a(t)^3},
\]
where a subscript 0 stands for evaluation at $t=0$. Denoting $\omega_0=3\,n_0a_0^3>0$, the equation for $\phi$ becomes
\begin{equation}\label{psieq}
\dot{\phi}(t)=-\frac{\omega_0\sigma}{a(t)^3}.
\end{equation}
Note that the equation for $\phi$ contains only matter terms evaluated at time $t=0$. This is a general consequence of the fact that $\phi$ satisfies the homogeneous wave equation~\eqref{waveq}. 

From now on I shall restrict the discussion to the spatially flat case $k=0$. In this case it is possible to introduce ``cartesian'' coordinates such that the metric takes the form
\[
ds^2=-dt^2+a(t)^2(dx^2+dy^2+dz^2)
\]
and $a_0=1$.
The Einstein equations are
\[
3\left(\frac{\dot{a}}{a}\right)^2-\phi=\rho,\quad-2\frac{\ddot{a}}{a}-\left(\frac{\dot{a}}{a}\right)^2+\phi=\mathscr{P}.
\]
It is convenient to rewrite them as
\begin{align}
&\dot{H}=\frac{1}{3}\phi-\frac{1}{6}(\rho+3\mathscr{P})-H^2,\label{eqH}\\
 &H^2:=\left(\frac{\dot{a}}{a}\right)^2=\frac{1}{3}(\rho+\phi),\label{hamiltonian}
\end{align}
where $H=\dot{a}/a$ is the Hubble function.

Initial data at $t=0$ for the system~\eqref{matter}--\eqref{hamiltonian} consist of $H_0=H(0)$, $\phi_0=\phi(0)$ and $0\leq F_0(q)=F(0,q)=f_0(p_1,p_2,p_3)\in C^2(\R^3)$, such that $\rho_0$ is bounded and the Hamiltonian constraint~\eqref{hamiltonian} is satisfied at $t=0$. It will be assumed that  $H_0>0$, since this is the most interesting case for the applications (the universe is currently expanding). It will also be assumed that $f_0$ does not vanish identically, since otherwise the only solution is vacuum. It follows from~\eqref{hamiltonian} that for admissible initial data the value of $\phi_0$ is determined by the initial data $H_0$ and $f_0$ through the Hamiltonian constraint
\begin{equation}\label{initialboundpsi}
\phi_0=3H_0^2-\rho_0.
\end{equation}
Finally, $\phi_0$ will be identified with the present observed value of the cosmological constant. In particular, the case $\phi_0\geq 0$ is the most interesting for the applications. Since it is an ubiquitous quantity in cosmology, I shall also use
\begin{equation}\label{omegazero}
\Omega_0=\frac{\rho_0}{3H_0^2}=1-\frac{\phi_0}{3H_0^2},
\end{equation}
together with the following additional Hubble-normalized constants:
\begin{equation}\label{nzero}
N_0=\frac{n_0}{3H_0^2},\quad \Sigma_0=\frac{\sigma}{H_0}.
\end{equation}

Using~\eqref{hamiltonian}, eq.~\eqref{eqH} becomes $\dot{H}=-(\rho+\mathscr{P})/2$; in particular
 $H$ is always decreasing and the following two possibilities may occur:
%By regular solution in the interval $(0,T)$ we mean that $0<a\in C^2((0,T))$, $H,\phi\in C^1((0,T))$, $0\leq F\in C^2((0,T)\times (0,\infty))$ (which implies $\rho>0$ on $(0,T)$). 
%In the remaining of this section the future behavior of solutions to the system~\eqref{matter}--\eqref{hamiltonian} will be analised, assuming that they have the required regularity to justify the following calculations.\footnote{A more mathematical analysis of the system~\eqref{Einstein}-\eqref{FokkerPlanck} will be presented elsewhere.}  
%There exist two possibilities:
\begin{itemize}
\item[1)] Either $H$ remains positive for all times, or
\item[2)] $H$ vanishes at some time $t_*>0$. 
\end{itemize}
In the first case the universe is expanding forever in the future, in the second case it collapses into a singularity in finite time. Which of the two possibilities occurs depends on the initial value of the cosmological scalar field $\phi$. If $\phi_0>0$ is such that $\phi$ remains always positive, then, by~\eqref{hamiltonian}, $H$ remains positive as well. On the other hand, if $\phi$ vanishes at some time $\bar{t}$, then by~\eqref{eqH}, $\dot{H}\leq -|\phi|/3 -H^2$, for $t>\bar{t}$ and therefore $H$ will become zero at some time $t_*>\bar{t}$. After this moment, the Hubble function continues to decrease until it diverges to $-\infty$ at some finite time $t_{**}>t_{*}$, and $a(t)\to 0$ as $t\to t_{**}$. 

Since $\phi$ is decreasing, future collapse in finite time occurs when $\phi_0\leq 0$. In particular, even though the initial value of $\phi_0$, i.e., the present observed cosmological constant, were vanishing, the model under study still predicts future collapse into a singularity in finite time. Even more remarkable is that the same conclusion holds if $\phi_0$ is positive and small. In fact it will now be shown that if
\begin{equation}\label{psizerocollapse}
0<\phi_0<\frac{n_0\sigma}{H_0},
\end{equation}
then $\phi$ becomes zero in finite time. To see this, suppose on the contrary that $\phi>0$ for all times. Since $H$ is decreasing we have
\[
\dot{a}(t)=H a(t)\leq H_0a(t).
\]
Gr\"onwall's inequality\footnote{See~\cite{evans}.} now gives
\begin{equation}\label{esta1}
a(t)\leq e^{H_0t}.
\end{equation}
Combining~\eqref{esta1} with~\eqref{psieq} we get
\[
\lim_{t\to +\infty}\phi(t)=\phi_0-\omega_0\sigma\int_0^{+\infty}\frac{ds}{a(s)^3}\leq\phi_0-\frac{n_0\sigma}{H_0},
\]
which is negative under assumption~\eqref{psizerocollapse}. Whence $\phi$ must vanish at some time $\bar{t}\in (0,+\infty)$, a contradiction to the hypothesis that $\phi$ was always positive and the claim is proved. Note also that by~\eqref{initialboundpsi}, and using the Hubble normalized variables~\eqref{omegazero}-\eqref{nzero}, 
the condition found for future collapse in finite time (including $\phi_0\leq 0$) can be rewritten as
\[
\Omega_0>1-N_0\Sigma_0,
\]
which is satisfied by all admissible initial data when the diffusion constant is sufficiently large:
\[
\Sigma_0\geq N_0^{-1}.
\] 
%In the standard, diffusion-free cosmological models, the fact that $\Omega_0>1$ is an indication that $\sigma=1$, i.e., that the universe is closed. This is not longer the case in the presence of diffusion, since recollapse can occur also in an open flat universe. 
For $\Sigma_0=0$, the model reduces to the Einstein-Vlasov system with cosmological constant $\Lambda=\phi=const$ and the condition for future collapse in finite time becomes $\Omega_0>1$, which is equivalent to $\Lambda<0$. The latter condition is optimal, since the Einstein-Vlasov system with non-negative cosmological constant in a flat Robertson-Walker geometry is forever expanding in the future for all initial data.
 
Next I will show the existence of initial data for the cosmological scalar field for which $\phi$, and therefore $H$, remains positive for all times, i.e., the universe is forever expanding in the future. By what has just been proved, this can only happen if $\sigma< 3H_0^3/n_0$, i.e., $\Sigma_0<N_0^{-1}$. The following argument requires a stronger bound on $\sigma$, precisely
\begin{equation}\label{asssigma}
\sigma< \frac{H_0^3}{n_0},\quad \text{i.e.,}\quad \Sigma_0<(3N_0)^{-1},
\end{equation} 
which will be therefore assumed from now on. Thanks to~\eqref{asssigma} we can choose 
\begin{equation}\label{phizeroglobal}
\phi_0\geq\frac{3n_0\sigma}{H_0},\quad \text{i.e.,}\quad \Omega_0\leq1-3N_0\Sigma_0.
\end{equation} 
%Precisely, I will prove that for 
%\begin{align}
%\phi_0\geq\overline{\phi_0}&=\min\left( 2\sqrt{3N_0\sigma H_0},3H_0^2+\frac{N_0\sigma}{H_0}\right)\nonumber\\
%&=\left\{\begin{array}{ll} 3H_0^2+\frac{N_0\sigma}{H_0} & \ \text{for $N_0\sigma \leq 3H_0^3$}\%\ 2\sqrt{3N_0\sigma H_0} & \ \text{for $N_0\sigma>3H_0^3$}\end{array}\right.\label{psizero}
%\end{align}
%$\phi$ remains positive for all times; in particular, for all $t>0$:
%\[
%\phi(t)\geq 3H_0^2,\ \text{if $N_0\sigma\geq 3H_0^3$},\quad\phi(t)\geq \sqrt{3N_0\sigma H_0}, \% \text{if $N_0\sigma<3H_0^3$.}
%\]
% 
By continuity, there exists $0<\bar{t}\leq +\infty$ such that $\phi(t)>0$ on $[0,\bar{t})$ and if $\bar{t}$ were finite we would have $\phi(\bar{t})=0$. Our goal is to prove that $\bar{t}=+\infty$. 
Using the bound $\mathscr{P}\leq \rho/3$ and that $H$ is decreasing on $[0,\bar{t})$, we obtain
\begin{align*}
\dot{H}&=\frac{1}{3}\phi-\frac{1}{6}(\rho+3\mathscr{P})-H^2\geq \frac{1}{3}\phi-\frac{1}{3}\rho-H^2\\
&=\frac{2}{3}\phi-2H^2\geq -2H^2.
\end{align*}
It follows that
\[
H(t)\geq \frac{H_0}{1+2H_0t}\Rightarrow a(t)\geq a_0(1+2H_0 t)^\frac{1}{2},
\]
for all $t\in [0,\bar{t})$.
Then from~\eqref{psieq} we obtain
\[
\phi(t)=\phi_0-\omega_0\sigma\int_0^t\frac{ds}{a(s)^3}>\phi_0-\frac{3n_0\sigma}{H_0}.
\]  
Assuming $\bar{t}<\infty$, assumption~\eqref{phizeroglobal} entails $\phi(\bar{t})>0$, a contradiction, and therefore  $\bar{t}=+\infty$ must hold.
Moreover, if the strict inequality holds in~\eqref{phizeroglobal} our model behaves like a perfect fluid Robertson-Walker model with negative spatial curvature ($k=-1$), since $H$ never vanishes, not even asymptotically.

To summarize:
\begin{itemize}
\item Future collapse in finite time takes place for all admissibile initial data when $\Sigma_0\geq N_0^{-1}$ and for $\Omega_0>1-N_0\Sigma_0$ when $\Sigma_0<N_0^{-1}$. 
\item Unlimited future expansion takes place for initial data such that $\Sigma_0<(3N_0)^{-1}$ and $\Omega_0\leq 1-3N_0\Sigma_0$. 
\end{itemize}
More accurate estimates than those presented in this section are required to cover the full range of initial data. Numerical simulations may also provide a useful insight to the problem.

\section{Summary and final remarks}
In this paper a new model to describe the dynamics of particles undergoing velocity diffusion in general relativity has been proposed. The main goal was to understand the influence of the diffusion forces on the large scale dynamics of the galaxies in the universe, in the approximation where each galaxy can be represented by a particle of the system.

The dynamics of the particle system was described through the kinetic distribution function $f$, which was assumed to satisfy a Fokker-Planck equation without friction on the tangent bundle of spacetime. In the absence of diffusion, the Fokker-Planck equation reduces to the Vlasov equation in general relativity. 

Since the energy momentum tensor associated to a solution of the Fokker-Planck equation is not divergence free, the coupling with the Einstein equations is inconsistent. Two solutions to this problem have been proposed: Either assume the existence of additional matter fields in spacetime, or modify the Einstein equations. This paper explored in some details the case in which a cosmological scalar field is added to the left hand side of the Einstein equations, providing a natural generalization of the cosmological constant theories. It has been shown that diffusion changes drastically the qualitative properties of even the simplest cosmological models, namely the flat Robertson-Walker spacetimes.
  
The paper cannot be concluded without a few comments on the physical limitations of the models presented here. First of all it should be emphasized (once again) that diffusion is not a fundamental interaction, but rather an approximation for the dynamics of two particle systems, one of which is dynamically dominant on the other.  While in the Newtonian and special relativistic diffusion theories the dynamics of the ``dominant component" can be broken apart from the dynamics of the other ``weak component", the latter being therefore relegated to a background thermal bath, in general relativity this approximation is consistent 
and some interaction between the two components must always be taken into account. The next best ``diffusion approximation" consists in assuming the simplest possible dynamics for the thermal bath, as the cosmological scalar field theory developed in this paper. However it seems that the only theoretical way to justify a specific model for the thermal bath is by deriving the diffusion approximation model from a (still unknown) fundamental theory for the interaction of the two particle systems.

A final comment on the diffusion models introduced in this paper is that they lack of an action principle formulation, which is again due to their phenomenological, rather than fundamental, nature.

\begin{center}
{\bf Acknowledgments}
\end{center}
This research was carried out while I was a long term visitor of the Center of Mathematics for the Applications (CMA) in Oslo. I would like to thank the members of the CMA and particularly Xavier Raynaud for several discussions on the topics of this paper. Finally I would like to thank an anonymous referee for her/his valuable comments and for pointing out ref.~\cite{kamiab} to me.

\end{document}